\documentclass[]{article}

\usepackage{color}
\usepackage{graphicx}
\usepackage{mathptmx}

\begin{document}
\noindent	
{\bf \Large Influence of inhibitory synapses on the criticality of excitable
	neuronal networks}\\
	
\noindent F S Borges\textsuperscript{1,*}, P R Protachevicz\textsuperscript{2,3},V
Santos \textsuperscript{2}, M S Santos\textsuperscript{3}, E C
Gabrick\textsuperscript{2}, K C Iarosz\textsuperscript{3,4,5,*}, E L
Lameu\textsuperscript{6}, M S Baptista\textsuperscript{7}, I L
Caldas\textsuperscript{3}, A M Batista\textsuperscript{2,3,8}\\
	
{\footnotesize \noindent 
\textsuperscript{1}Center for Mathematics, Computation, and	Cognition, Federal University of ABC, 09606-045, S\~ao Bernardo do Campo, SP,	Brazil. \textsuperscript{2}Graduate Program in Science - Physics, State	University of Ponta Grossa, 84030-900, Ponta Grossa, PR, Brazil. 
\textsuperscript{3}Institute of Physics, University of S\~ao Paulo,	05508-900, S\~ao Paulo, SP, Brazil. 
\textsuperscript{4}Faculdade de Tel\^emaco Borba, 84266-010, Tel\^emaco Borba, PR, Brazil. 
\textsuperscript{5}Graduate Program in Chemical Engineering, Federal University of Technology Paran\'a, 84016-210, Ponta Grossa, PR, Brazil. \textsuperscript{6} Cell Biology and Anatomy Department, University of Calgary, AB T2N 4N1, Calgary, AB, Canada. \textsuperscript{7} Institute for Complex Systems and Mathematical Biology, University of Aberdeen, AB24 3UE, Aberdeen, Scotland, UK. \textsuperscript{8}Department of Mathematics and Statistics, State	University of Ponta Grossa, 84030-900, Ponta Grossa, PR, Brazil.\\
}
\footnotesize	Corresponding author: fernandodasilvaborges@gmail.com, kiarosz@gmail.com.
		
\begin{abstract}  
In this work, we study the dynamic range of a neuronal network of excitable 
neurons with excitatory and inhibitory synapses. We obtain an analytical 
expression for the critical point as a function of the excitatory and inhibitory
synaptic intensities. We also determine an analytical expression that gives 
the critical point value in which the maximal dynamic range occurs. Depending
on the mean connection degree and coupl\-ing weights, the critical points can
exhibit ceasing or ceaseless dynamics. However, the dynamic range is equal in
both cases. We observe that the external stimulus mask some effects of
self-sustained activity (ceaseless dynamic) in the region where the dynamic
range is calculated. In these regions, the firing rate is the same for
ceaseless dynamics and ceasing activity. Furthermore, we verify that excitatory
and inhibitory inputs are approximately equal for a network with a large number
of connections, showing excitatory-inhibitory balance as reported
experimentally.


\end{abstract}

	\normalsize
\section{Introduction}

The relation between stimuli and sensation is one of the main research topics
in Psychophysics \cite{chescheider2013}. Stimulus of different sources and
intensities can cause different responses in the sensory system
\cite{Stevens1975}. In the early 19th century, Weber and Fechner proposed that
stimuli-respon\-se relation correspond to a logarithmic function
\cite{kinouchi06,Levina2007}.  In the 1950s, Stevens proposed that
stimuli-response relation is given by a power law \cite{stevens08}. Due to
physiological and anatomical limitation, the relation between stimuli and
response have upper and lower limits. The stimuli difference, between the
smaller and bigger sensation, define the dynamic range (DR) associated with its
sense \cite{murray93}. In the context of biological systems, e.g. neuronal
networks, the DR corresponds to the ability to differentiate the intensity of
external stimulus
\cite{gollo09}.

The DR is proportional to the logarithm of the ratio between the largest value
of the external applied stimulus in which the response is close to saturation
of the firing rate and the smallest value of the external applied stimulus
in which it is weak to modify the firing rate. The human sense of sight
can perceive changes in about ten decades of luminosity and the hearing covers 
twelve decades in a range of intensities of sound pressures
\cite{stevens08,chialvo06}. The DR of the human vision plays an important role
in the design of display devices \cite{reinhard10}, where as the
hearing case it is relevant to cochlear implants \cite{spahr07}. 

The DR of a neuronal network increases with the network size until it reaches a
saturation value \cite{batista14}. The increase of the DR value is also
associated with the increase of the number of excitatory chemical synapses
\cite{viana14,iarosz2012}. Borges et al. \cite{borges} reported the
complementary effect of chemical and electrical synapses on the enhance of the
DR. Protachevicz et al. shown that chemical synapses can enhance DR 
of the neural network submitted to external stimuli \cite{Protachevicz2018b}. 

The mammalian brain is composed of excitatory and inhibitory neurons 
\cite{adini1997}. The balance between excitation and inhibition plays a crucial
role in the transmission of information, signal propagation, and regular firing
of neurons in many brain areas \cite{kandel00,buzsaki06}. Neuronal networks
with excitatory and inhibitory neurons have been considered to describe the 
dynamics of primary visual cortex \cite{adini1997,kurant06}, cortical firing
patterns \cite{Borges2017,Prota2018,Borges2020,protachevicz19}, and synaptic 
plasticity mechanisms \cite{Borges2017b,Borges2017c,Lameu2018,Lameu2018b}.

Kinouchi and Copelli \cite{kinouchi06} proposed a model of an excitable network
based on Erd\"os-R\'enyi (ER) random gra\-phs \cite{erdos59}. They
demonstrated that the DR is maximised at the critical point of a non
equilibrium phase transition. A theoretical approach to study the effects of
network topology on the DR was presented by Larremore et al.
\cite{Larremore2011,Larremore2011b}, in which was considered only excitatory
nodes. Pei et al. \cite{pei2012} investigated the collective dynamics of
excitatory-inhibito\-ry excitable networks in response to external stimuli.
They found that the DR is maximised at the critical point of phase transition
which depends only on the excitatory connections.  

The spiking dynamics of a network of excitable excitatory nodes resulting from
an initial stimulus ceases after a typically short time at a critical point
\cite{kinouchi06}. However, when inhibitory nodes are considered the collective
dynamic can become self-sustaining as shown by Larre\-more et al.
\cite{Larremore2014}. They showed this behaviour considering an additive
probabilistic model, where excitatory nodes increase the probability of
activation of their nei\-ghbours, and inhibitory nodes decrease the probability.
In addition, at critical point the collective dynamics can become
self-sustainable (ceaseless dynamic) if a fraction of inhibitory nodes is
greater than a threshold. However, in their model they did not consider a
refractory period and, for this reason, the neuronal firing rate obtained is
higher than the experimentally observed. When refractoriness is included in the
model, it is possible to obtain the critical point leading to realistic firing
patterns \cite{copeli2019,mauricio2019}.

In this work, we investigate the criticality and dynamic range of a cellular
automaton modelling a neuronal network in which the neurons are connected by
means of excitatory and inhibitory chemical synapses \cite{viana14,borges}. In
order to understand the relationship between maximisation of the DR and the
critical self-sustainable activity, we consider a refractory period in the model
like the one proposed by Larremore et al. \cite{Larremore2014}. With the
refractory period, the model exhibits more realistic firing rates and critical
self-sustained activity. In our simulations, we observe a transition from
ceaseless dynamics to ceasing activity when the mean connection degree of the
network is increased. We observe that the external stimulus mask effects of
self-sustained activity in the region where the DR is calculated, and the
firing rate is the same for the ceaseless dynamics and ceasing activity.
Furthermore, we obtain an analytical expression for the DR as a function of the
mean excitatory and inhibitory synaptic intensities. In a network with a large
number of connections, we show that the maximal DR value occurs in the critical
points where excitatory and inhibitory inputs are approximately equal. In this
situation, the neuronal network is in a balanced state. Shew et al. \cite{shew}
showed experimentally that the DR is maximised when the excitatory and
inhibitory synaptic input are balanced. Our work thus provides theoretical
explanations for this experimental result. 

The paper is organised as follows. In Section $2$, we introduce the model.
Section $3$ presents our analytical results about the dynamic range. In the
last Section, we draw our conclusions.

\section{Model}

\begin{figure}[htbp!]
	\centering
	\includegraphics[scale=0.5]{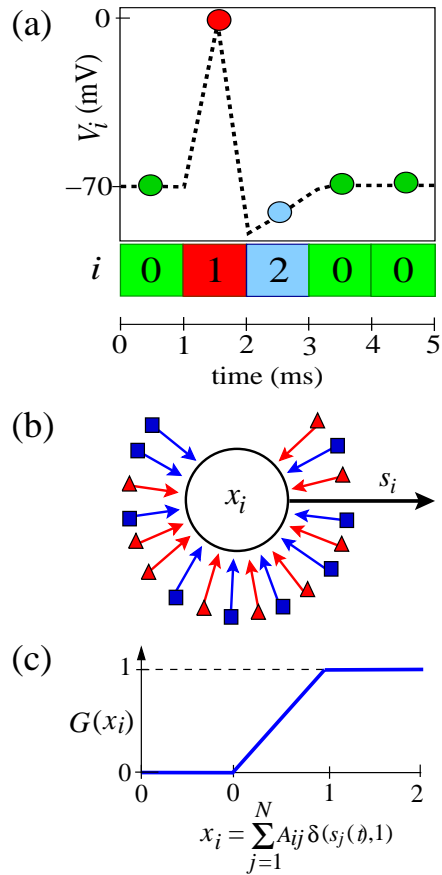}
	\caption{Representation of the neuronal activity by a cellular automaton with
		$n=3$ states. (a) Illustrative membrane potential for each neuron $i$, where
		$s_i$ represent the rest ($s_i=0$), the active ($s_i=1$), and refractory 
		($s_i=2$) states. (b) Chemical synaptic inputs arriving in the neuron $i$. Red
		triangles and blue squares represent the excitatory and inhibitory inputs,
		respectively. (c) The neuronal activation probability ($G(x_i)$) is given by
		a function of all chemical inputs arriving in the neuron $i$ at time $t$.}
	\label{Fig1}
\end{figure}

We consider a $n$ states cellular automaton model composed of $N$ excitable
elements. The state of each neuron $i$ is described by the variable $s_i$
($i=1,..., \ldots n$). In this representation, each neuronal state is
associated with the neuronal activity \cite{kinouchi06,copelli02}. The resting
state is given by $s_i=0$, the excited state by $s_i=1$, and $s_i = 2,...,n-1$ 
are the refractory states. The elements can not be excited during the
refractory states. In the model, we consider excitatory and inhibitory neurons 
\cite{Larremore2014}. Inhibitory and excitatory inputs are related to the
excitatory and inhibitory neurons, respectively. To model the interaction of
the synaptic inputs, we considere a probability function. The activation
probability of a node in the resting state is given by function 
$G (x_i)$ \cite{Larremore2014}
\begin{equation}
G(x_i) = G \left(\sum_{j=1}^{N} A_{ij} \;\delta (s_j (t),1)\right),
\end{equation}
where $G(x_i)=0$ for $x_i \leq 0$, $G(x_i)=x_i$ for $0 < x_i < 1$, and
$G(x_i)=1$ for $x_i \geq 1$. $G(x_i)$ is a piecewise linear function, known as
transfer function, with three pieces. The weighted matrix $\mathbf{A}$ has
elements $A_{ij}>0$ for excitatory connections and $A_{ij}<0$ for inhibitory
connections. The Kronecker delta $\delta(a,b)$ is equal to $1$ when $a=b$ and
zero otherwise. The dynamics of both excited and the refractory states are
deterministic. If $s_i=1$, in the next time steps the state is updated to
$s_i=2$, and so forth, until $s_i=n-1$, returning to the resting state $s_i=0$
in the next time step. The fractions of excitatory and inhibitory nodes
correspond to $f_{\rm ex}$ and $f_{\rm in}$, respectively, and the condition
$f_{\rm ex}+f_{\rm in}=1$ is always satisfied. In order to simplify the analysis,
we arrange the $i$ indexes as $1\leq i\leq f_{\rm ex}N$ for excitatory nodes,
whereas $f_{\rm ex}N+1\leq i\leq N$ for inhibitory ones. Fig. \ref{Fig1} displays
an schematic illustration of (a) the neuronal dynamics for $n=3$ states, (b) a
neuron receiving chemical synaptic inputs and (c) the function $G(x_i)$ as a
function of the sum of all synaptic inputs.

The neuronal response at a given time $t$ can be qu\-antified using the density
of spiking neurons
\begin{equation}
p(t)=\frac{1}{N}\sum_{i=1}^{N}\delta(s_i(t),1),
\end{equation}
which is interpreted as the probability for a random neuron to be in the
excited state at time $t$. With the time series of $p(t)$,  we calculate the
average firing rate
\begin{equation}
F=\overline {p(t)}=\frac{1}{T}\sum_{t=1}^{T}p(t),
\end{equation}
where $T$ is the time window chosen to calculate the average. 

In this work, we consider random networks and for this case the update
equations are the same for both excitatory and inhibitory nodes \cite{pei2012}.
Our networks are built according to the Erd\"os-R\'enyi random graphs with
probability equal to $K/(N-1)$, where $K$ is the average degree of connections
of the network. Assuming that the events of the neighbours of an excited node
are statistically independent for large $t$, we obtain the following mean field
map for the density of spiking neurons
\begin{equation}\label{mapp}
p(t+1)= [1-(n-1)p(t)] (\eta+G(x)-\eta G(x)),
\end{equation}
where the external stimulus $\eta=1-\exp{(-r\Delta t)}$ is a Poisson process
with mean perturbation rate $r$ in the time interval $\Delta t$
\cite{kinouchi06}. In our simulations, we use $\Delta t=1$.

Setting the weights $A_{ij}=S_{\rm ex}$ for the excitatory connections and
$A_{ij}=-S_{\rm in}$ for the inhibitory ones, when the network reaches a
stationary state, the mean value of $x_i$ is given by 
\begin{equation}
\langle x\rangle=f_{\rm ex}KS_{\rm ex}p(t)-f_{\rm in}KS_{\rm in}p(t).
\end{equation}
Defining $\sigma_{\rm ex}=KS_{\rm ex}$ and $\sigma_{\rm in}=KS_{\rm in}$, we obtain
\begin{equation}
\langle x\rangle=(f_{\rm ex}\; \sigma_{\rm ex}-f_{\rm in} \; \sigma_{\rm in}) \;p (t).
\end{equation}
In the stationary state we have $p(t+1)=p(t)=p^*$ and $F \approx p^*$.
Substituting in Eq. (\ref{mapp}), and considering the case of no external
perturbation ($\eta = 0$), we get
\begin{equation}
F_0=(1-(n-1)F_0) \; G(x).
\end{equation}
In the regime $0<(f_{\rm ex} \; \sigma_{\rm ex}-f_{\rm in} \; \sigma_{\rm in})F<1$,
the model implies $G(x) = x$, and therefore
\begin{equation}
F_0=(1-(n-1)F_0) \; (f_{\rm ex}\; \sigma_{\rm ex}-f_{\rm in} \; \sigma_{\rm in})F_0.
\end{equation}
Solving for $F_0$ we get
\begin{equation}\label{F0}
F_0=\frac{1-(f_{\rm ex}\sigma_{\rm ex}-f_{\rm in}\sigma_{\rm in})^{-1}}{n-1}.
\end{equation}
There is a phase transition from ceasing activity ($F_0 = 0$) to ceaseless
activity ($F_0 > 0$). In the critical point of this phase transition
($F_0\rightarrow 0$), we obtain  
\begin{equation}\label{sigmac}
\sigma_{\rm in}=\frac{f_{\rm ex}\sigma_{\rm ex}-1}{f_{\rm in}}.
\end{equation}
This relation shows that the critical point in the model is given by
$f_{\rm ex}\sigma_{\rm ex}\geq 1$, implying the necessity of a minimum fraction
of excitatory neurons. In addition, we observe that
$f_{\rm ex}\sigma_{\rm ex}\approx f_{\rm in}\sigma_{\rm in}$ for
$\sigma_{\rm ex} \gg 1$. Then, for a highly connected network
($\sigma\propto K$), we obtain approximately the same amount of excitatory and
inhibitory mean inputs from probabilistic synapses. In this situation, our model
exhibits a state which is critical and balanced.

\begin{figure}[htbp!]
	\centering
	\includegraphics[scale=0.28]{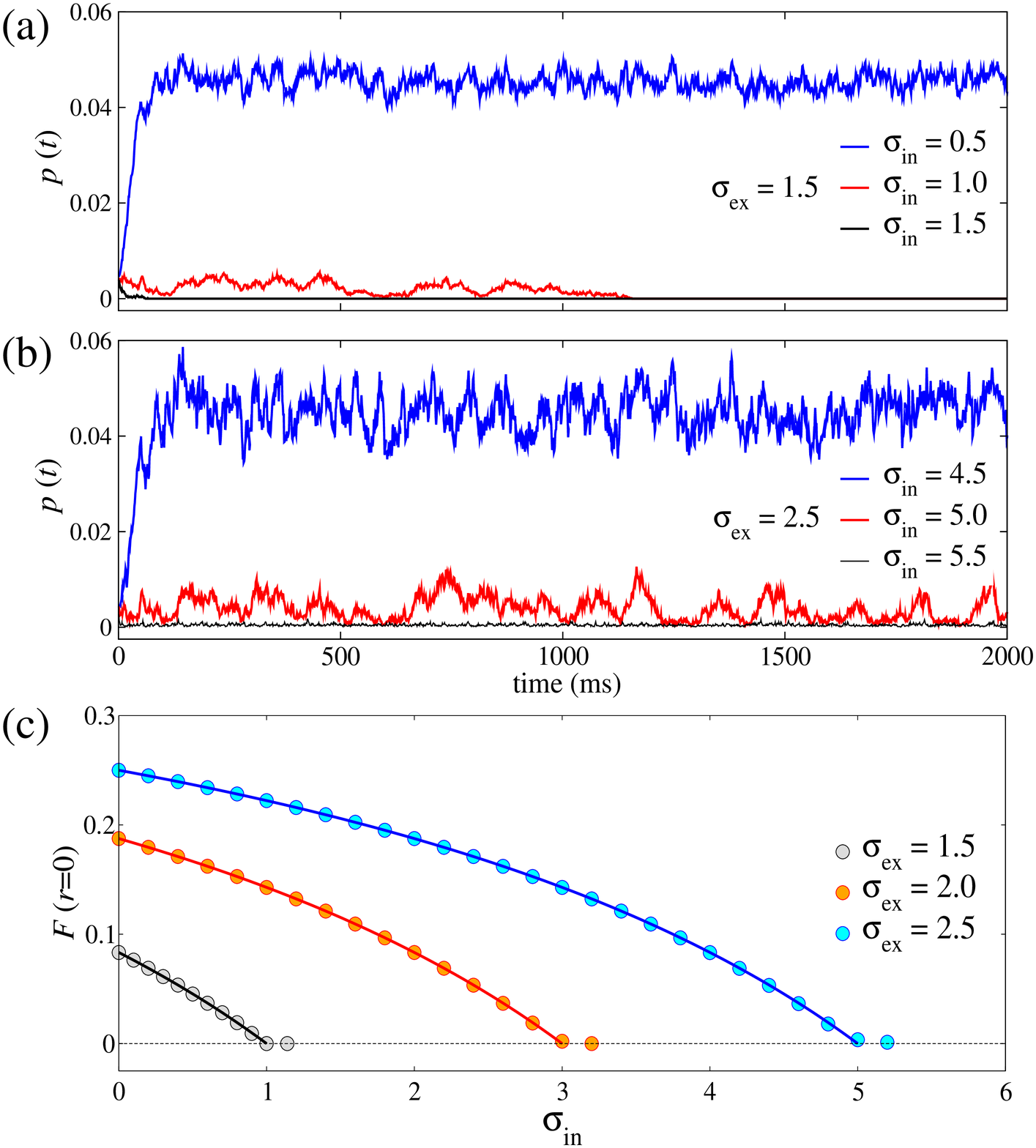}
	\caption{Time series of the density of spiking neurons for subcritical (black
		line), critical (red line), and supercritical (blue line) values of
		$\sigma _{\rm in}$ for (a) $\sigma_{\rm ex}=1.5$ and (b) $\sigma_{\rm ex}=2.5$. In
		(c), we plot the  average firing rate as a function of $\sigma_{\rm in}$ for
		$\sigma_{\rm ex}=1.5$ (black circles), $\sigma_{\rm ex}=2.0$ (red circles) and
		$\sigma_{\rm ex}=2.5$ (blue circles). The points are obtained from numerical
		simulations while the curves are given by Eq. \ref{F0}. The parameters are
		$N=10^5$, $K=10^4$, $r=0$, $n=3$, and $f_{\rm ex}=0.8$.}
	\label{Fig2}
\end{figure}

In this work, we split in three theoretical firing regi\-mes that depende of Eq. \ref{F0} and
of the parameters $f_{\rm ex}$, $\sigma_{\rm ex}$, $f_{\rm in}$, and $\sigma_{\rm in}$.
(i) if $F_0 < 0$ we have a subcritical regime; (ii) if $F_0 = 0$ we have a critical regime; and 
(iii) if $F_0 > 0$ we have a supercritical regime.
In Figs. \ref{Fig2}(a) and \ref{Fig2}(b), we show the density of spiking 
neurons without external perturbation as a function of the time for values of
$\sigma_{\rm in}$ in the subcritical, critical, and supercritical regime for
different values of $\sigma_{\rm ex}$. We choose randomly $0.4 \%$ of neurons 
to be active ($s_i=1$) at $t=0$. In Fig. \ref{Fig2}(c), we show the
relation between $F_0$ and $\sigma_{\rm in}$ for some values of $\sigma_{\rm ex}$.
We verify that the theoretical results given by Eq. \ref{F0} are in agreement
with our numerical simulations.

\begin{figure}[htbp!]
	\centering
	\includegraphics[scale=0.37]{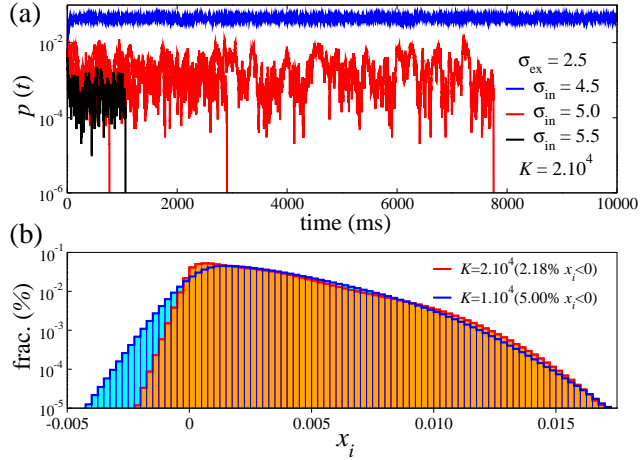}
	\caption{(a) Time series of the density of spiking neurons for the subcritical 
		(black line), critical (red line), and supercritical (blue line) values of
		$\sigma _{\rm in}$ with $\sigma_{\rm ex}=2.5$. (b) Distribution of $x_i$ values for
		the average degree of connections $K=2\times 10^4$ (red) and $K=1\times 10^4$
		(blue). Parameters are $N=10^5$, $f_{\rm ex}=0.8$, $\sigma_{\rm ex}=1.5$, and
		$\sigma_{\rm ex}=2.5$.}
	\label{Fig3}
\end{figure}

When a great number of neurons presents $x_i<0$, even at the critical point,
in the numerical simulations $F$ can be positive for a large time span. In Fig. \ref{Fig2}, we see that
the spiking activity ceases rapidly when $\sigma_{\rm ex}=1.5$, whereas it is
persistent at the critical point when $\sigma_{\rm ex}=2.5$. We verify that the
activity is not persistent if we increase the average degree of connections.
Fig. \ref{Fig3}(a) exhibits the density of spiking neurons considering 
$K=2\times 10^4$, for subcritical, critical (three different initial conditions) 
and supercritical values of $\sigma_{\rm in}$. In Fig. \ref{Fig3}(b), we plot the 
distribution of $x_i$ for $1000$ time steps,
$N=10^5$, $K=10^4$ (blue), and $K=2\times 10^4$ (red). In both cases, we find
$\langle x\rangle$ and $F\approx 0.0031$. In the first case, approximately
$2.18 \%$ of $x_i$ present negative values. In the second case, about $5.00\%$
of $x_i$ are less than zero. We observe that greater values of
$S_{\rm ex}=\sigma_{\rm ex}/K$ and $S_{\rm in}=\sigma_{\rm in}/K$ contribute for the
persistent activity at the critical point.

\section{Dynamic Range (DR)}

\begin{figure}[htbp!]
	\centering
	\includegraphics[scale=0.29]{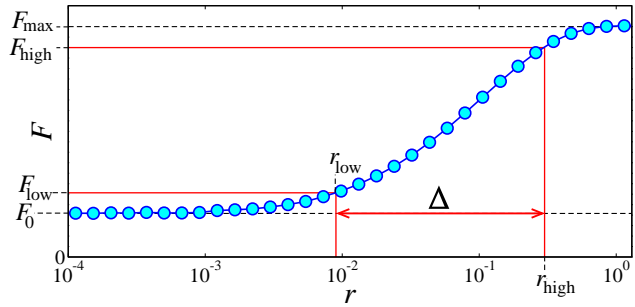}
	\caption{Mean firing rate as a function of intensity stimuli.}
	\label{Fig4}
\end{figure}

The behaviour of the average firing rate ($F$) as a function of the external
stimulus ($r$) shows a minimum and a maximum saturation ($F_0$ and $F_{\rm max}$,
respectively) for a range of $r$ values, as shown in Fig. \ref{Fig4}. The
DR is defined as
\begin{equation} \label{DR}
\Delta=10\log_{10}\frac{r_{\rm high}}{r_{\rm low}},
\end{equation}
where $\Delta$ is the stimulus interval (measured in dB) in which changes in $r$ can
be perceived as changes in $F$, and it is between the disregarding stimuli that
cause a response small to be distinguished from $F_0$ and the saturation
$F_{\rm max}$ \cite{kinouchi06}. The interval [$r_{\rm low}$,$r_{\rm high}$] is found
from its correspondent in $F$, [$F_{\rm low}$,$F_{\rm high}$], where
$F_{\rm high}=F_0+0.95(F_{\rm max}-F_0)$ and $F_{\rm low}=F_0+0.05(F_{\rm max}-F_0)$.	

For $0<(f_{\rm ex}\; \sigma_{\rm ex}-f_{\rm in} \; \sigma_{\rm in})F<1$ (in the
stationary state) we approximate Eq. \ref{mapp} as
\begin{eqnarray}
F&=&[1-(n-1)F]  [\eta + (f_{\rm ex}\; \sigma_{\rm ex}-f_{\rm min} \; \sigma_{\rm in})F \nonumber \\
&-& (f_{\rm ex}\; \sigma_{\rm ex}-f_{\rm in} \; \sigma_{\rm in})\eta F]. 
\label{eq13}
\end{eqnarray}
Rearranging the terms, we obtain
\begin{eqnarray}
& &\left[(n-1)(f_{\rm ex}\; \sigma_{\rm ex}-f_{\rm min} \; \sigma_{\rm in}) (1-\eta)\right]F^2 \nonumber \\
&+&\left[1+(n-1)\eta-(f_{\rm ex}\; \sigma_{\rm ex}-f_{\rm min} \; \sigma_{\rm in})(1-\eta)\right]F \nonumber \\
&-& \eta= 0.  \label{eq14}
\end{eqnarray}
As $\eta$ depends on $r$, by solving Eq. \ref{eq14}, we are able to determine
the dependence of the average firing rate on the mean perturbation rate
$r$, as well as its dependence on all the parameters of the network.

\begin{figure}[htbp]
	\centering
	\includegraphics[scale=0.19]{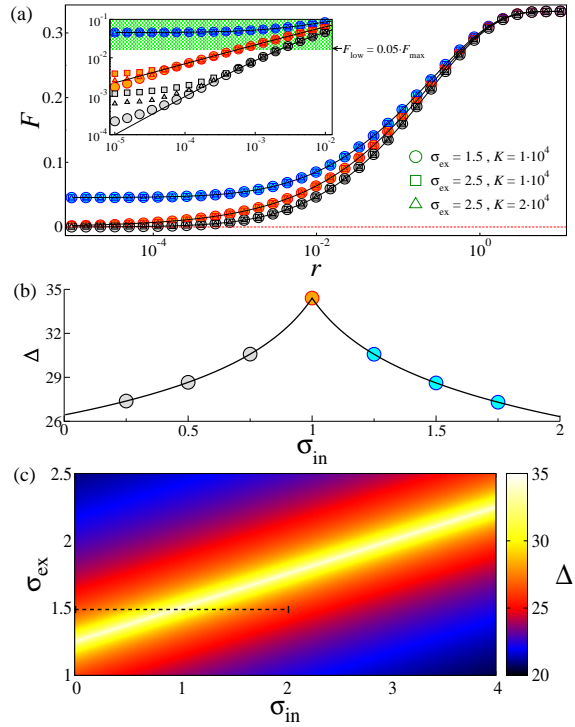}
	\caption{(a) Average firing rate ($F$) as a function of the mean perturbation
		rate ($r$). The black, red, and blue symbols correspond to subcritical,
		critical and supercritical values of $\sigma _{\rm in}$, respectively. (b)
		Dynamic range as a function of $\sigma_{\rm in}$ for $\sigma_{\rm ex}=1.5$. The
		coloured circles are obtained by means of simulations and the black lines
		represent the theoretical results from the analytical expression. Dynamic
		range of Fig. (b) is indicated on the $\sigma_{\rm ex} \times \sigma_{\rm in}$
		parameter space by a dash line in Fig. (c). We consider $N=10^5$, $K=10^4$, and
		$f_{\rm ex}=0.8$.}
	\label{Fig5}
\end{figure}

In Fig. \ref{Fig5}(a), we plot $F$ as a function of $r$ for subcritical,
critical and supercritical values of $\sigma_{\rm in}$. The lines represent the
theoretical values from the solution of expression \ref{eq14} and the symbols
are obtained through numerical simulations. In the inset of Fig. \ref{Fig5}(a), we
show a magnification to demonstrate that there are diferences between the
theoretical and the numerical values of $F$ for $r$ values out of the region
where DR is calculated (green).

For a cellular automaton with $n$ states, the maximum average firing rate is
given by $F_{\rm max}=1/n$. Deriving $F_0$ in Eq. \ref{F0}, $F_{\rm low}$ and
$F_{\rm high}$ can be obtained. Then, $\eta _{\rm low}$ and $\eta _{\rm high}$ can be
calculated directly by
\begin{equation} \label{eta_lh}
\eta _{\rm low,high}=\frac{\lambda F_{\rm low,high}}{1-\lambda F_{\rm low,high}} 
\left[\frac{1}{\lambda-(n-1)\lambda F_{\rm low,high}}-1\right],
\end{equation}
where we substitute $\lambda=f_{\rm ex}\sigma_{\rm ex}-f_{\rm in}\sigma_{\rm in}$
for convenien\-ce. Now we calculate $r_{\rm low}$ and $r_{\rm high}$ according to
\begin{equation} \label{r_lh}
r_{\rm low,high}=-\ln |1-\eta_{\rm low,high}|.
\end{equation}
Using the equations \ref{F0}, \ref{eta_lh}, \ref{r_lh}, and the
expressions for $F_{\rm max}$, $F_{\rm low}$, and $F_{\rm high}$, we calculate the
dynamic ran\-ge.

In Fig. \ref{Fig5}(b), we compare our numerical and the theoretical results.
We verify that the maximum DR occurs for $\sigma _{\rm in}=1$, which is the
critical point for the considered parameters ($\sigma_{\rm ex}=1.5$). In Fig.
\ref{Fig5}(c), the color scale represents the value of DR for each pair
$(\sigma _{\rm ex},\sigma _{\rm in})$. The dashed line indicates the range taken
in (b). From the figure, we see that the maximum value of DR follows the line
given by the critical point expression $\sigma_{\rm in}=4\sigma_{\rm ex}-5$ 
(Eq. \ref{sigmac}). Since
the excitatory-inhibitory ratio is $4$, the mean input approaches zero as 
the $\sigma_{\rm ex}$ value increases. Therefore, the model shows both the
critical and balanced state in a network with a great number of connections,
where the weights are not small. For instance, for $K=2 \times 10^4$ and
$S_{\rm ex}=0.5$, we obtain $\sigma_{\rm ex}=KS_{\rm ex}=10^4$ and the critical
$\sigma_{\rm in}=4\times 10^4-5$. The ration between excitatory and inhibitory
input is $\frac{4\times\sigma_{\rm ex}}{\sigma_{\rm in}}\approx 1.0013$. In this
situation, the DR is maximum and closes to a balanced state.


\section{Conclusions}

The firing dynamics of a network of excitable excitatory nodes resulting from an
initial stimulus ceases after a typically short time at a critical point.
However, when inhibitory nodes are considered the collective dynamic can become
self-sustained. In this work, we build a cellular automaton model with
excitatory and inhibito\-ry connections. In our network, we consider that the
connections have different weights. We find an expression that relates the mean
of excitatory and inhibitory weights at the critical point. We also calculate
an expression for the dynamic range and show that at the critical point it
reaches its maximal value.

Depending of mean connection degree and coupl\-ing weights, the critical points
can exhibit ceasing or ceaseless dynamics (self-sustained activity). However,
the dynamic range is equal in both cases. We observe that the external stimulus
mask some effects of self-sustained activity in the region where the DR is
calculated. In these regions, the firing rate is the same for ceaseless
dynamics and ceasing activity. Furthermore, we show that at the critical point
the amount of excitatory and inhibitory inputs can be approximately equal in a
densely connected network. This result showing excitatory-inhibitory balanced
was  experimentally reported by Shew et al. \cite{shew}.

In future works, we plan to consider other network topologies, such as
small-world and scale-free, to study the influence of inhibitory synapses on
the criticality of excitable neuronal networks.


\section*{Acknowledgment}
This study was possible by partial financial support fr\-om the following
Brazilian government agencies: Fun\-da\c c\~ao Arauc\'aria, National Council
for Scientific and Technological Development, Coordination for the Improvement
of Higher Education Personnel, and S\~ao Pa\-ulo Research Foundation
(2015/07311-7, 2017/18977-1, 2018/03211-6, 2020/04624-2).

\end{document}